\documentclass[11pt,draftclsnofoot,onecolumn]{IEEEtran}
\pdfoutput=1
\usepackage[left=1.25in,right=1.25in,top=1in,bottom=1in]{geometry}
\usepackage{epsfig,citesort,cite,multirow}
\usepackage{graphicx,setspace}
\usepackage{url}
\usepackage{algorithm}
\usepackage{algorithmic}
\usepackage{url}

\usepackage{booktabs}

\usepackage{amsmath,amssymb,amsfonts}
\usepackage{enumerate}
\usepackage{epsfig}
\usepackage{amsxtra}
\usepackage{psfrag}
\usepackage{color}
\usepackage[usenames,dvipsnames]{xcolor}

\usepackage[colon-equals]{mathtools}
\usepackage{microtype} 
\usepackage{xcolor}
 \colorlet{dblue}{blue!80!black}
\usepackage[citecolor=dblue,colorlinks=true, filecolor=dblue,linkcolor=dblue,menucolor=dblue,urlcolor=dblue]{hyperref}

\floatstyle{plaintop}
\newfloat{recipe}{thp}{lor}
\floatname{recipe}{Recipe}

\newcommand{\qtq}[1]{\quad\text{#1}\quad}

\newcommand{\vct}[1]{\mathbold{#1}}
\newcommand{\mtx}[1]{\mathbold{#1}}
\newcommand{\Proj}{\ensuremath{\mtx{\Pi}}} 
\newcommand{\Desc}{\mathcal{D}} \newcommand{\sdim}{\delta}

\newtheorem{definition}{Definition}
\newtheorem{theorem}{Theorem}

\newcommand{\bitem}{\begin{itemize}}
\newcommand{\eitem}{\end{itemize}}

\newcommand{\beqn}{\begin{equation}}
\newcommand{\eeqn}{\end{equation}}
\newcommand{\balign}{\begin{align}}
\newcommand{\ealign}{\end{align}}

\usepackage{amssymb,amsmath}
\usepackage{subfig,graphicx,colortbl,multicol}
\usepackage{fixmath}
\usepackage{tikz}
\usepackage{subfig}
\usepackage{cases}
\usepackage[numbers]{natbib}

\usepackage{graphicx}
\usepackage{multirow}
\usepackage{url}

\renewcommand{\phi}{\varphi}

\renewcommand{\mid}{\mathrel{\mathop{:}}}

\newcommand{\subjectto}{\quad\text{\textnormal{subject to}}\quad}

 \newcommand{\zerovct}{\mathbf{0}} 
\newcommand{\Id}{\mathbf{I}}

\providecommand{\mathbbm}{\mathbb} 
\newcommand{\R}{\mathbbm{R}}

\newcommand{\argmin}{\operatorname*{arg\; min}}

\newcommand{\Expect}{\operatorname{\mathbb{E}}}

\newcommand{\normal}{\textsc{Normal}}

\newcommand{\hvct}[1]{\widehat{\vct{#1}}}

\newcommand{\transp}{t}

\newcommand{\norm}[1]{\left\Vert {#1} \right\Vert}

\newcommand{\lone}[1]{\norm{#1}_{{\ell_1}\!}}

\newcommand{\enorm}[1]{\norm{#1}_2}

\DeclareMathOperator{\conv}{conv}

\begin{document}
\title{Convexity in source separation: \\ Models, geometry, and algorithms}
\author{\IEEEauthorblockN{Michael B. McCoy, Volkan Cevher, Quoc Tran Dinh, \\   Afsaneh Asaei, and Luca Baldassarre }\thanks{The authors thank Joel A. Tropp for his helpful and detailed comments on this work. MBM is supported by ONR awards N00014-08-1-0883 and N00014-11-1002, AFOSR award
FA9550-09-1-064. Work of VC, QTD, and LB is supported in part by the European Commission under Grant MIRG-268398, ERC Future Proof, SNF 200021-132548, SNF 200021-146750 and SNF CRSII2-147633. The work of AA is funded by SNF NCCR IM2.}}
{\singlespace
\maketitle}

\makeatletter{}Source separation or \emph{demixing} is the process of extracting multiple components entangled within a signal. Contemporary signal processing  presents a host of difficult source separation problems, from interference cancellation to background subtraction,  blind deconvolution, and even dictionary learning. Despite the recent progress in each of these  applications, advances in high-throughput sensor technology place demixing algorithms under pressure to accommodate extremely high-dimensional signals,  separate an ever larger number of sources, and cope with more  sophisticated signal and mixing models. These difficulties are exacerbated by the need for real-time action in automated decision-making systems. 

Recent advances in convex optimization provide a simple framework for efficiently solving numerous difficult demixing problems.  This article provides an  overview of the emerging field, explains the theory that governs the underlying procedures, and surveys  algorithms that solve them efficiently. We aim to equip practitioners with a toolkit for constructing their own demixing algorithms that \emph{work}, as well as concrete intuition for \emph{why} they work.

\vspace{-3mm}
\subsection*{Fundamentals of demixing}\vspace{-1mm}

The most basic model for mixed signals is a \emph{superposition model}, where we observe a mixed  signal \(\vct z_0\in \R^d\) of the form
\begin{equation}\label{eq:signal-model-intro}
 \vct z_0 = \vct x_0 + \vct y_0,
\end{equation}
and we wish to determine the component signals \(\vct x_0\) and \(\vct y_0\).  This simple model appears in many guises.  Sometimes, superimposed signals come from basic laws of nature.  The amplitudes of electromagnetic waves, for example, sum together at a receiver, making the superposition model~\eqref{eq:signal-model-intro} common in wireless communications. Similarly, the additivity of sound waves makes superposition models natural in speech and audio processing.  

Other times, a superposition  provides a useful, if not literally true, model for more complicated nonlinear phenomena. Images, for example, can be modeled as the sum of constituent features---think of stars and galaxies that sum to create an image of a piece of the night sky~\cite{StaMurFad:10}. In machine learning, superpositions can describe hidden structure~\cite{ChaSanPar:11}, while in statistics, superpositions can model gross corruptions to data~\cite{CanLiMa:11}.  These models also appear in texture repair~\cite{liang2012repairing}, graph clustering~\cite{CheJalSan:13}, and line-spectral estimation~\cite{Bhaskar13}.  
 
 A conceptual understanding of demixing in all of these applications rests on two key ideas.
\begin{LaTeXdescription}
\item[Low-dimensional structures:] Natural signals in high dimensions often cluster around low-dimensional structures with few degrees of freedom relative to the ambient dimension~\cite{baraniuk2010low}. Examples include bandlimited signals, array observations from seismic sources, and natural images.  By identifying the  convex functions that encourage these low-dimensional structures, we can derive convex programs that  disentangle structured components from a signal.
\item[Incoherence:] Effective demixing requires more than just structure. To distinguish multiple elements in a signal, the components must look different from one another.  We capture this idea by saying that two structured families of signal are \emph{incoherent} if their constituents appear very different from each other.  While demixing is impossible without incoherence, sufficient  incoherence typically leads to provably correct demixing procedures.
\end{LaTeXdescription}
The two notions of structure and incoherence above also appear at the core of recent developments in information extraction from incomplete data in compressive sensing and other linear inverse problems~\cite{CanWak:08,chandrasekaran2012convex}. The theory of demixing extends these ideas to a richer class of signal models, and it leads to a more coherent theory of convex methods in signal processing.

While this article primarily focuses on mixed signals drawn from the superposition model~\eqref{eq:signal-model-intro}, recent extensions to  \emph{nonlinear} mixing models arise in blind deconvolution, source separation,  and nonnegative matrix factorization~\cite{ahmed2012blind,saunderson2012diagonal,BitReRec:12}. We will see that  the same techniques that let us demix superimposed signals reappear in  nonlinear demixing problems. 

\vspace{-2mm}
\subsection*{The role of convexity}

Convex optimization provides a unifying theme for all of the demixing problems discussed above. This framework is based on the idea that many structured signals possess corresponding convex functions that encourage this structure~\cite{chandrasekaran2012convex}.  By combining these functions in a sensible way, we can develop convex optimization procedures that demix a given observation.  The geometry of these functions lets us understand when it is possible to demix a superimposed observation with incoherent components~\cite{McCTro:12}.  The resulting convex optimization procedures usually have both theoretical and practical guarantees of correctness and computational efficiency.

To illustrate these ideas, we consider a classical but surprisingly common demixing problem: separating impulsive signals from sinusoidal signals, called the \emph{spikes and sines} model.  This model appears in many applications, including star--galaxy separation in astronomy, interference cancellation in communications, inpainting and speech enhancement in signal processing~\cite{DonHuo:01,StaMurFad:10}. 

While individual applications feature additional structural assumptions on the signals, a simple low-dimensional signal model effectively captures the main idea present in all of these works: \emph{sparsity}.  A vector \(\vct x_0 \in \R^d\) is \emph{sparse} if most of its entries are equal to zero. Similarly, a vector \(\vct y_0\in \R^d\) is \emph{sparse-in-frequency} if its discrete cosine transform (DCT) \(\mtx D \vct y_0\) is sparse, where \(\mtx D\in \R^{d\times d}\) is the matrix that encodes the DCT.  Sparse vectors capture impulsive signals like pops in audio, while sparse-in-frequency vectors explain smooth objects like natural images. Clearly, such signals look different from one another. In fact, an arbitrary collection of spikes and sines is linearly independent or \emph{incoherent} provided that the collection is not too big~\cite{DonHuo:01}.

\begin{figure}[t]
  \centering
    \begin{tabular}{c}
      \includegraphics[width=.92\columnwidth]{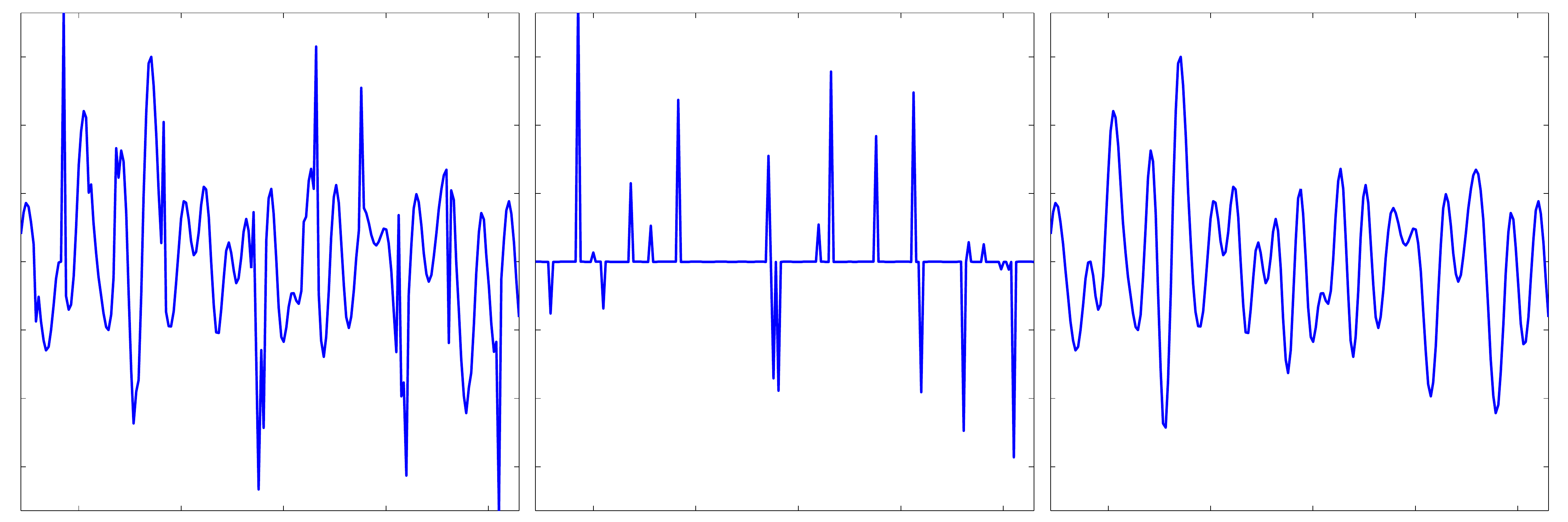} \\
        \includegraphics[width=0.9\columnwidth]{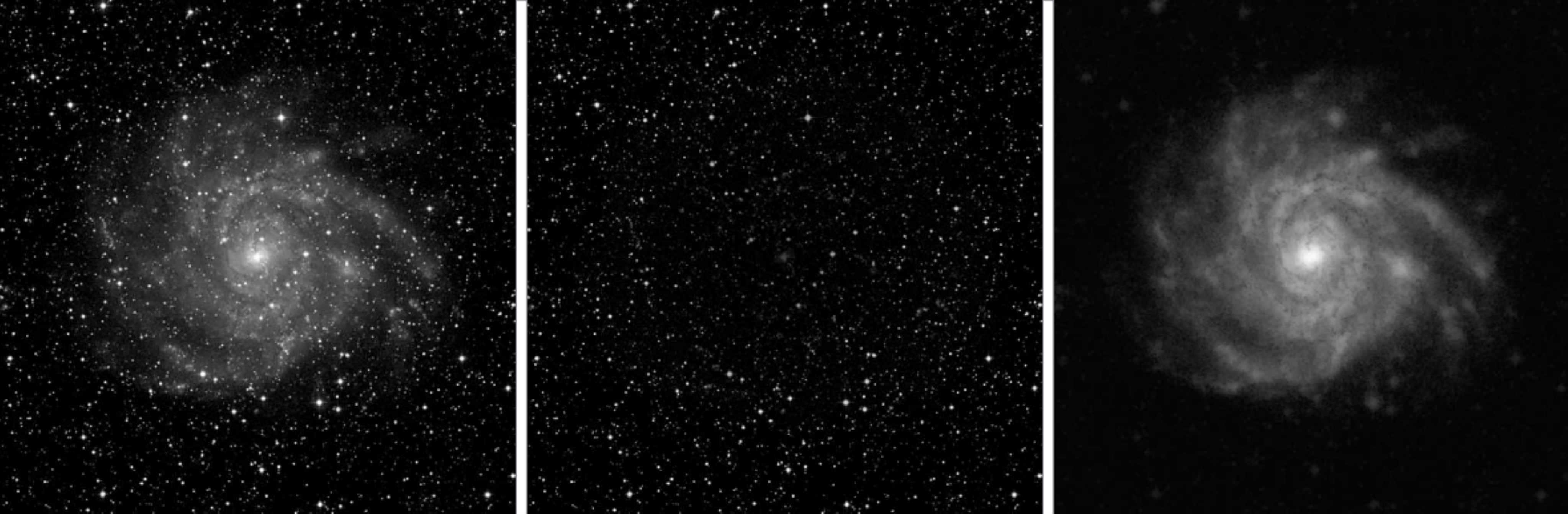} 
    \end{tabular}
    \vspace{-4mm}
    \begin{flushleft}
      \hspace{6.75mm}\tiny{\textnormal{Image credit: NASA}}
    \end{flushleft}
    \vspace{-4mm}
  \begin{tabular}{ccc}
    \hspace{8mm} Observation \(\vct z_0\)   & \hspace{11mm} Sparse component \(\vct x_0\) &  \hspace{2mm} DCT-sparse component \(\vct y_0\)
  \end{tabular}
\caption{\it [Top] A perfect separation of spikes from sinusoids from their additive mixture with \eqref{eq:ell1}. The original signal (left) is perfectly separated into its sparse component (center) and its DCT-sparse component (right) [Bottom] Star-galaxy separation using~\eqref{eq:ell1} on a real astronomical image.  The original (left) is separated into a  starfield (center) corresponding to a nearly sparse component and a galaxy (right) corresponding to a nearly DCT-sparse component. }   \label{fig:svg}
  \vspace{-6mm}
\end{figure}

Is it possible to demix a superimposition \(\vct z_0= \vct x_0 + \vct y_0\) of spikes and cosines into its constituents?  One approach is to search for the \emph{sparsest possible} constituents that generate the observation \(\vct z_0\):
\begin{equation}\label{eq:ell0}
    \left[ \check{\vct x}, \check{\vct y} \hspace{.3mm}\right] := \argmin_{\vct x, \vct y \in \R^n} \bigl \{\|\vct x \|_{0}+ \lambda \| \mtx D\vct y \|_{0}: \vct z_0 =  \vct x +  \vct y \bigr\},
\end{equation}
where the $\ell_0$ ``norm'' measures the sparsity of its input, and  $\lambda>0$ is a regularization parameter that trades the relative sparsity of solutions. Unfortunately, solving \eqref{eq:ell0} involves an intractable computational problem.  However, if we replace the \(\ell_0\) penalty with the convex $\ell_1$-norm, we arrive at a classical sparse approximation program~\cite{DonHuo:01}:
\begin{equation}\label{eq:ell1}
    [ \hvct x,\hvct y\,] := \argmin_{\vct x, \vct y \in \R^n} \bigl \{\|\vct x \|_{1}+ \lambda \| \mtx D\vct y \|_{1}: \vct z_0 =  \vct x +  \vct y \bigr\}.
\end{equation}
This key change to the combinatorial proposal~\eqref{eq:ell0} offers numerous benefits.  First, the procedure~\eqref{eq:ell1} is a convex program, and a number of  highly efficient algorithms are available for its solution. Second, this procedure admits provable guarantees of correctness and noise-stability under incoherence.  Finally, the demixing procedure~\eqref{eq:ell1} often performs admirably in practice.  

Figure~\ref{fig:svg} illustrates the performance of~\eqref{eq:ell1} on both a synthetic signal drawn from the spikes-and-sines model above, as well as on a real astronomical image.  The resulting performance for the basic model is quite appealing even for real data that mildly violates the modeling assumptions. Last but not least, this strong baseline performance can be obtained in fractions of seconds with simple and efficient algorithms.

\vspace{-3mm}
\subsection*{Outline}

The combination of efficient algorithms, rigorous theory, and impressive real-world performance are a hallmark of the convex demixing paradigm described in this article. Below, we provide a unified treatment of demixing problems using convex geometry and optimization starting with Section~\ref{sec:rolling-your-own}. Section~\ref{sec:geometry-demixing} describes some emerging connections between statistics and geometry that characterizes the success and the failure of convex demixing. Section \ref{sec:numerics:-burn-it} describes scalable algorithms for practical demixing. Sections~ \ref{sec:examples} and \ref{sec:new-direct-nonl} trace the recent frontier in source separation. We not only ground the new theory on compelling signal processing applications but also point out how we can tackle \emph{nonlinear} demixing problems.

\makeatletter{}
\section{Demixing made easy}
\label{sec:rolling-your-own}

This section provides a recipe to generate a convex program that accepts a mixed signal \(\vct z_0 = \vct x_0 + \vct y_0\) and returns a set of demixed components.  
The approach requires two ingredients.  First, we must identify convex functions that promote the structure we expect in \(\vct x_0\) and \(\vct y_0\).   Second, we combine these functions together into a  convex objective.   This simple and versatile approach easily extends to multiple signal components and undersampled observations.
\vspace{-3mm}
\subsection*{Structure-inducing  convex functions}
We say that a signal has structure when it has  fewer degrees of freedom than the ambient space. Familiar examples of structured objects include sparse vectors, sign vectors, and low-rank matrices.  It turns out that each of these structured families have an associated convex function, called an atomic gauge, adapted to their specific features~\cite{chandrasekaran2012convex}.

The general principle is simple.  Given a set of \emph{atoms} \(\mathcal{A} \subset \R^d\), we say that a signal $\vct x \in \R^d$ is \emph{atomic} if it is formed by a  sum of a  small number of scaled atoms. For example, sparse vectors are atomic relative to the set of standard basis vectors because every sparse vector is the sum of just a few standard basis vectors.  For a  more sophisticated example,  recall that the singular value decomposition implies that low-rank matrices are the sum of a few rank-one matrices. Hence, low-rank matrices are atomic relative to the set \(\mathcal{A}\) of  all rank-one matrices.

\begin{figure}
  \begin{center}
        \includegraphics[width=0.3\columnwidth]    {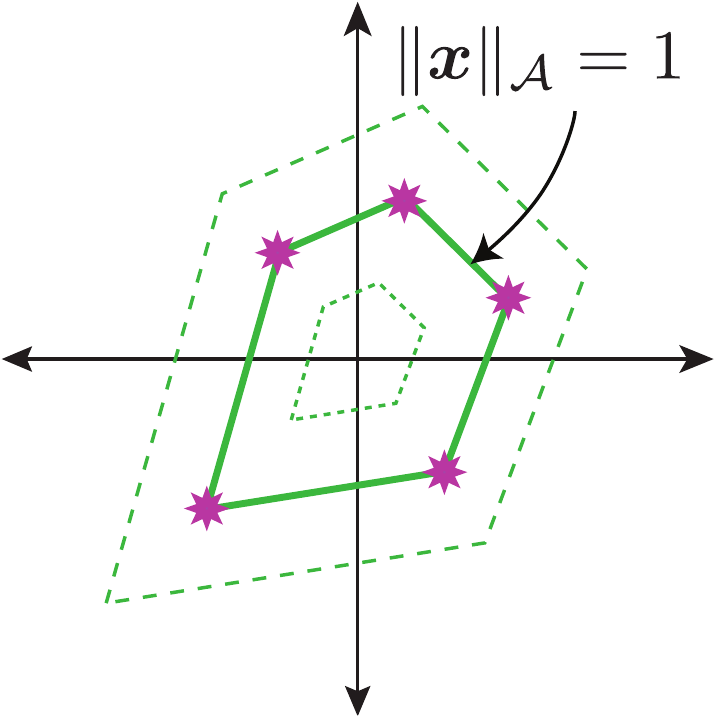}    \hspace{2cm}
    \includegraphics[width=0.3\columnwidth]    {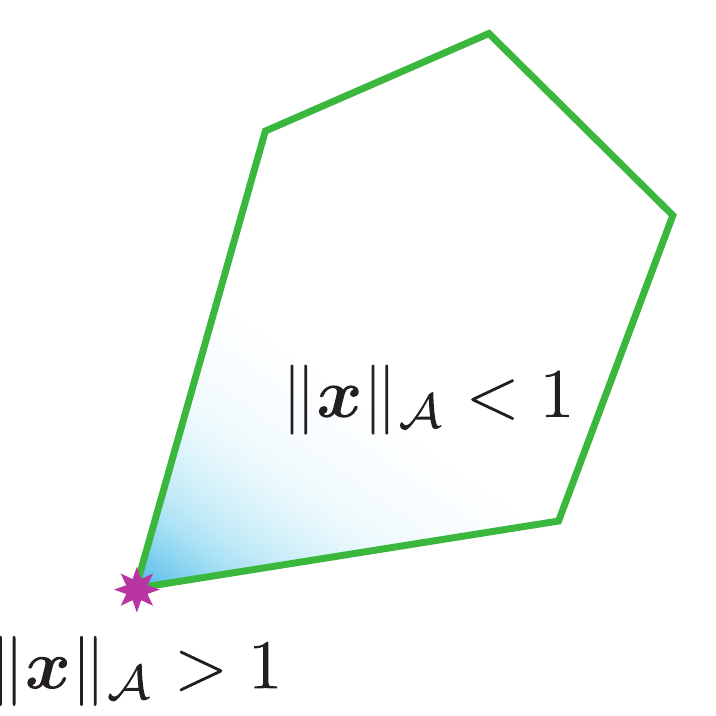}
  \end{center}
  \vspace{-12pt}
  \caption[An atomic gauge.]{\it {\sl [Left]} An atomic set \(\mathcal{A}\) consisting of five atoms {\it(stars)}.  The ``unit ball'' of the atomic gauge \(\|\cdot\|_{\mathcal{A}}\) is the closed convex hull of \(\mathcal{A}\) {\sl (heavy line)}.  Other level sets {\it (dashed lines)} of the gauge are dilations of the unit ball. {\sl [Right]}  At an atom {\sl(star)}, the unit ball of \(\|\cdot\|_{\mathcal{A}}\) tends to have sharp corners.  Most perturbations away from this atom increase the value of \(\|\cdot\|_\mathcal{A}\), so the atomic gauge often penalizes complex signals that are comprised of a large number of atoms. }
  \label{fig:atomic-gauge}
  \vspace{-5mm}
\end{figure}

We can define  a  function that measures the inherent complexity of signals relative to a given set \(\mathcal{A}\). One natural measure is the \emph{fewest} number of scaled atoms required to write a signal using atoms from \(\mathcal{A}\), but unfortunately, computing this quantity can be computationally intractable. Instead, we define the \emph{atomic gauge}  \(\|\vct x\|_{\mathcal{A}}\) of a signal \(\vct x\in \R^d\) by
\begin{equation*}
    \|\vct x\|_{\mathcal{A}} := \inf \big \{\lambda > 0 \mid \vct x \in \lambda\cdot \conv(\mathcal{A}) \big \},
  \end{equation*}
  where \(\conv(\mathcal{A})\) is the convex hull of \(\mathcal{A}\).  In other words,
the level sets of the atomic gauge are the scaled versions of the convex hull of all the atoms \(\mathcal{A}\) (Figure~\ref{fig:atomic-gauge} [Left]).

By construction, atomic gauges are ``pointy'' at atomic vectors. This property means that most deviations away from the atoms result in a rapid increase in the value of the gauge, so that the function tends to penalize deviations away from simple signals (Figure~\ref{fig:atomic-gauge} [Right]).  The pointy geometry plays an important role in the theoretical understanding of demixing, as we will see in Section~\ref{sec:geometry-demixing}. 

\begin{table}
\caption{\label{tab:atomic_norms}{Example signal structures and their atomic gauges~\cite{CheDonSau:98,chandrasekaran2012convex}.}  \it The top two rows correspond to vectors while the bottom  three refer to matrices.  The vector norms extend to matrix norms by treating \(m\times n\) matrices as length-\(mn\)  vectors.  The expression \(\enorm{\vct x}\) denotes the Euclidean norm of the vector \(\vct x\), while  \(\sigma_i(\mtx X)\) returns the \(i\)th singular value of the matrix \(\mtx X\).} 
\vspace{-12pt}
\begin{center}
  \small
    \begin{tabular}{ccc}
      \toprule
      {\bf Structure} & {\bf Atomic set} & {\bf Atomic gauge \(\|\cdot\|_\mathcal{A}\)} \\
      \midrule
      Sparse vector & Signed basis vectors \( \{\pm \vct e_i\}\)
      &
      \begin{tabular}{c}
        \(\ell_1\) norm \\ \(\|\vct x\|_{\ell_1}=\sum_{i} |x_i|\) 
      \end{tabular}\\[10pt]
      \begin{tabular}{c}
        Binary \\ sign vector
      \end{tabular}
      & Sign vectors \(\{\pm 1\}^d\) &
      \begin{tabular}{c}
        \(\ell_\infty\) norm \\
        \(\|\vct x\|_{\ell_\infty} = \max_i |x_i|\)
      \end{tabular}
                              \\
      \midrule
      Low-rank matrix&
      \begin{tabular}{c}
        Rank-\(1\) matrices \\ \(\{\vct u \vct v^t\colon\|\vct u \vct v^t\|_{F}=1\}\)
      \end{tabular}
      &
      \begin{tabular}{c}
        Schatten  \(1\)-norm          \\
        \(\|\mtx X\|_{S_1}=\sum_{i} \sigma_i(\mtx X)\)
      \end{tabular}
      \\[10pt]
      Orthogonal matrix &
      \begin{tabular}{c}
        Orthogonal matrices \\
        \(\{\mtx O\colon \mtx O \mtx O^t = \mathbf{I}\}\)
      \end{tabular}
      &
      \begin{tabular}{c}
        Schatten \(\infty\)-norm \\
        \(\|\mtx X\|_{S_\infty} = \sigma_1(\mtx X)\)
      \end{tabular}
      \\[10pt]
      \begin{tabular}{c}
        Row-sparse\\  matrix
      \end{tabular}
      &
      \begin{tabular}{c}
        Matrices w/one nonzero row \\
        \(\{\vct e_i \vct v^\transp\mid \enorm{\mtx v}= 1\}\)
      \end{tabular}
      &
      \begin{tabular}{c}
        Row-\(\ell_1\) norm \\
        $\|\mtx X\|_{\ell_1/\ell_2} $
      \end{tabular}
      \\
      \bottomrule
    \end{tabular}
\end{center}
\vspace{-24pt}
\end{table}

A number of common structured families and their associated gauge functions appear in Table~\ref{tab:atomic_norms}. More sophisticated examples include gauges for probability measures, cut matrices, and low-rank tensors. We caution, however, that not every  atomic gauge is easy to compute, and so we must take care in order to develop \emph{tractable} forms of atomic gauges~\cite{chandrasekaran2012convex,bach2010structured}. Surprisingly, it is sometimes easier to compute the value of atomic gauges than it is to compute the (possibly nonunique) decomposition of a vector into its atoms~\cite{BitReRec:12}.  We will return to the discussion of tractable gauges when we discuss numerical schemes further in Section~\ref{sec:numerics:-burn-it}.

\vspace{-4mm}
 \subsection*{The basic demixing program} \label{sec:comb-ingr}

Suppose that we know the signal components \(\vct x_0\) and \(\vct y_0\) are atomic with respect to the known atomic sets \(\mathcal{A}_x\) and \(\mathcal{A}_y\).    In this section, we describe how to use the atomic gauge functions \(\norm{\cdot}_{\mathcal{A}_x}\) and \(\norm{\cdot}_{\mathcal{A}_y}\) defined  above to help us demix the components \(\vct x_0\) and \(\vct y_0\) from the observation \(\vct z_0\). 

Our intuition developed above indicates that the values \(\|\vct x_0\|_{\mathcal{A}_x}\) and \(\| \vct y_0\|_{\mathcal{A}_y}\) are relatively small because the vectors \(\vct x_0\) and \(\vct y_0\) are atomic with respect to the atomic sets \(\mathcal{A}_x\) and \(\mathcal{A}_y\).  This suggests that we search for constituents that generate the observation \emph{and} have small atomic gauges.  That is, we determine the demixed constituents \(\hvct x,\hvct y\)  by solving
\begin{equation}
  \label{eq:demix}
   [\hvct x,\hvct y ] =:\argmin_{\vct x,\vct y\in \R^d} \bigl\{ \|\vct x\|_{\mathcal{A}_x} + \lambda \|\vct y\|_{\mathcal{A}_y} \mid \vct x+\vct y = \vct z_0 \bigr\}.
\end{equation}
The  parameter \(\lambda>0\) negotiates a tradeoff between the relative importance of the atomic gauges, and the constraint \(\vct x+\vct y = \vct z_0\) ensures that our estimates \(\hvct x\) and \(\hvct y\) satisfy the observation model~\eqref{eq:signal-model-intro}.   The hope, of course, is that \(\hvct x = \vct x_0\) and \(\hvct y = \vct y_0\), so that the demixing program~\eqref{eq:demix} actually identifies the true components in the observation \(\vct z_0\).  

The demixing program~\eqref{eq:demix} is closely related to linear inverse problems and compressive sampling (CS)~\cite{CanWak:08,chandrasekaran2012convex}.  Indeed, the summation map \((\vct x,\vct y) \mapsto \vct x +\vct y\) is a linear operator, so demixing amounts to inverting an underdetermined linear system using structural assumptions.  The main conceptual difference between demixing and standard CS is that demixing treats the components \(\vct x_0\) and \(\vct y_0\) as unrelated  structures.  Also, unlike conventional CS, demixing does not require exact knowledge of the atomic decomposition, but only the value of the gauge.

The only link between the structures that appears in our recipe comes through the choice of tuning parameter \(\lambda\) in \eqref{eq:demix}, which makes these convex demixing procedures easily adaptable to new problems. In general, determining an optimal value of \(\lambda\)  may involve fine tuning or cross-validation, which can be quite computationally demanding in practice. Some theoretical guidance on the explicit choices regularization appears, for example, in~\cite{FoyMac:13,ChaSanPar:11,CanLiMa:11}.

\vspace{-4mm}
\subsection*{Extensions}
\label{sec:extensions-1}

There are many extensions of the linear superposition model~\eqref{eq:signal-model-intro}. In some applications, we are confronted with a  signal that is only partially observed---\emph{compressive} demixing. In others, we might consider an observation with additive noise, for instance, or a signal with more than two components.  The same ingredients that we introduced above can be used to demix signals from these more elaborate models. 

 For example, if we only see \(\vct z_0 = \mtx \Phi (\vct x_0 + \vct y_0)\), a linear mapping of the superposition, then we simply update the consistency constraint in the usual demixing program~\eqref{eq:demix} and solve instead
\begin{equation}\label{eq:gen1_demix}
   [\hvct x,\hvct y ] =:\argmin_{\vct x,\vct y\in \R^d} \bigl\{ \|\vct x\|_{\mathcal{A}_x} + \lambda \|\vct y\|_{\mathcal{A}_y} \mid \mtx \Phi(\vct x+\vct y) = \vct z_0\bigr\}.
\end{equation}
Some applications for this undersampled demixing model appear in image alignment~\cite{PenGanWri:12}, robust statistics~\cite{CheJalSan:13}, and graph clustering~\cite{CheJalSan:11}.  

Another straightforward extension involves demixing more than two signals. For example, if we observe \(\vct z_0 = \vct x_0 + \vct y_0 + \vct w_0\), the sum of three structured components, we can determine the components by solving
\begin{equation}\label{eq:gen2_demix}
  [\hvct x,\hvct y,\hvct w] := \argmin_{\vct x,\vct y,\vct w \in \R^d} \bigl\{\|\vct x\|_{\mathcal{A}_x} +\lambda_1 \|\vct y\|_{\mathcal{A}_y} +\lambda_2 \|\vct w \|_{\mathcal{A}_w} \mid \vct x +\vct y +\vct w = \vct z_0 \bigr\},
\end{equation}
where \(\mathcal{A}_w\) is an atomic set tuned to \(\vct w_0\), and as before, the parameters \(\lambda_i>0\) trade off the relative importance of the regularizers. This model appears, for example, in image processing applications where multiple basis representations, such as curvelets, ridgelets, shearlets, etc., explain different morphological components~\cite{StaMurFad:10}.   Further modifications along the lines above extend the demixing framework to a massive number of problems relevant to modern signal processing.

\makeatletter{}
\section{Geometry of demixing} 
\label{sec:geometry-demixing}
A critical question we can ask about a demixing program is ``When does it work?''  Answers to this question can be found by studying the underlying geometry of convex demixing programs.  Surprisingly, we can characterize the success and failure of convex demixing \emph{precisely} by leveraging a basic randomized model for incoherence. Indeed, the geometric viewpoint reveals a tight characterization of the success and failure of demixing in terms of geometric parameters that act as the ``degrees-of-freedom'' of the mixed signal.  The consequences for demixing are intuitive: demixing succeeds if and only if the dimensionality of the observation exceeds the total degrees-of-freedom in the signal.

\vspace{-8pt}
\subsection*{Descent cones and the statistical dimension}
\begin{figure}[t]
  \centering
  \includegraphics[width=0.65\columnwidth]{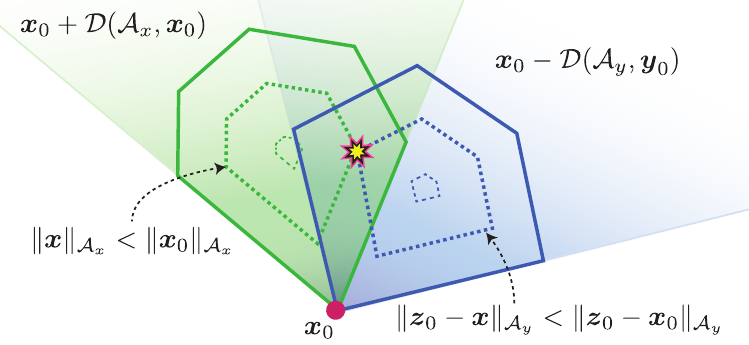}
  \caption{{Geometric characterization of demixing.} \it When the descent cones \(\Desc(\mathcal{A}_x,\vct x_0)\) and \(\Desc(\mathcal{A}_y,\vct y_0)\) share a line, then there is an optimal point \(\hvct x\) (star) for the demixing program~\eqref{eq:demix} not equal to \(\vct x_0\).  Conversely, demixing can succeed for some value of \(\lambda >0\) if  the two descent cones touch only at the origin. In other words, demixing can succeed if and only if \(\Desc(\mathcal{A}_x,\vct x_0) \cap -\Desc(\mathcal{A}_y,\vct y_0)=\{\zerovct\}\)~\cite{McCTro:12}.}
  \label{fig:success-demix}
\vspace{-6mm}
\end{figure}

Our study of demixing begins with a basic object that encodes the local geometry of a convex function.    The \emph{descent cone}  \(\Desc( \mathcal{A},\vct x)\) at a point \(\vct x\) with respect to an atomic set \(\mathcal{A}\subset \R^d\) consists of the directions where the gauge function \(\|\cdot\|_{\mathcal{A}}\) does not increase near  \(\vct x\).  Mathematically, the descent cone is given by
\begin{equation*}
  \Desc(\mathcal{A},\vct x):= \bigl\{\vct h \mid \|\vct x+\tau \vct h\|_{\mathcal{A}} \le \|\vct x\|_{\mathcal{A}} \;\;\text{for some}\;\; \tau >0\bigr\}. 
\end{equation*}
The descent cone encodes detailed information about the \emph{local} behavior of the atomic gauge \(\|\cdot\|_{\mathcal{A}}\) near  \(\vct x\). Since local optimality implies global optimality in convex optimization,  we can characterize when demixing succeeds in terms of a configuration of descent cones.  See Figure~\ref{fig:success-demix} for a precise description of this optimality condition.

In order to understand when the geometric optimality condition is likely to hold, we need a  measure for the ``size'' of cones.  The most apparent measure of size is perhaps the solid angle, which quantifies the amount of space occupied by a cone. The solid angle, however, proves inadequate for describing the intersection of cones even in the simple case of linear subspaces.  Indeed, linear subspaces are cones that take up no space at all, but when their dimensions are large enough, any two subspaces will always intersect along a line. Imagine trying to arrange two flat sheets of paper so that they only touch at their centers: impossible!   

It turns out that we find a much more informative statistic for demixing when we measure the proportion of space \emph{near} a cone, rather than the proportion of space \emph{inside} the cone.
 \begin{definition}
  Let \(C\subset \R^d\) be a closed convex cone, and denote by \(\Proj_C(\vct x):= \argmin_{\vct y \in C} \mbox{$\|\vct x-\vct y\|$}\) the closest point in \(C\) to \(\vct x\). We define the \emph{statistical dimension} \(\sdim(C)\) of a convex cone \(C\subset \R^d\)  by
  \begin{equation}\label{eq:sdim-def}
    \sdim(C) := \Expect \enorm{\Proj_C(\vct g)}^2,
  \end{equation}
  where \(\vct g \sim \normal(\zerovct, \Id)\) is a standard Gaussian random variable and the letter \(\Expect\) denotes the expected value.\end{definition}

The statistical dimension gets its name because it  extends many properties of the usual dimension of linear subspaces to convex cones~\cite{AmeLotMcC:13}, and it is closely related to the Gaussian width used in~\cite{chandrasekaran2012convex}.  Our interest here, however, comes from the interpretation of the statistical dimension as a  ``size'' of a cone.   A large statistical dimension \(\sdim(C)\approx d\) means that  \(\enorm{\Proj_C(\vct x)}^2\) is large  for most \(\vct x\in \R^d\)---that is, most points lie near the cone. On the other hand, a small statistical dimension  implies that most points lie far from~\(C\). We will see below that the statistical dimension  of descent cones provides the key parameter for understanding the success and failure of demixing procedures. 
Of course, a parameter is only useful if we can compute it.  Fortunately, the statistical dimension of descent cones is often easy to compute or approximate.  Several ready-made statistical dimension formulas and a step-by-step recipe for accurately deriving new formulas appear in~\cite{AmeLotMcC:13}. Some  useful approximate statistical dimension calculations can also be found in the works~\cite{chandrasekaran2012convex,FoyMac:13}.  As an added bonus, recent work indicates that  statistical dimension calculations are closely related to the problem of finding optimal regularization parameters~\cite[Thm.~2]{FoyMac:13}.

\vspace{-3mm}
\subsection*{Phase transitions in convex demixing}
The true power of the statistical dimension comes from its ability to predict \emph{phase transitions} in demixing programs. By phase transition, we mean the peculiar behavior where demixing programs switch from near-certain failure to near-certain success within a narrow range of model parameters. While the optimality condition from Figure~\ref{fig:success-demix} characterizes the success and failure of demixing, but it is often difficult to certify directly.  To understand how demixing operates in \emph{typical} situations, we need an incoherence model. 
One proposal to model incoherence assumes that the structured signals are oriented generically relative to one another.  This is achieved, for example, by assuming that the structured components are drawn structured relative to a rotated atomic set \(\mtx Q \mathcal{A}\), where \(\mtx Q\in \R^{d\times d}\) is a random orthogonal matrix~\cite{McCTro:12}. Surprisingly, this basic randomized model of incoherence leads to a rich theory with precise guarantees and predict {typical} behaviors well, and complements other phase transition characterizations in linear inverse problems~\cite{DonTan:10a,BayLelMon:12}.  Many works propose alternative incoherence models applicable to specific cases,  including~\cite{chandrasekaran2012convex,CanLiMa:11}, but these specific choices do not possess known phase transitions.  Under the random model of~\cite{McCTro:12}, however, a very general theory  is available.

\begin{theorem}[\cite{AmeLotMcC:13}] \label{thm:phase-trans}
Suppose that the atomic set of \(\vct x_0\) is \emph{randomly rotated}, i.e., that  \(\mathcal{A}_x = \mtx Q \tilde{\mathcal{A}}_x\) for some random rotation \(\mtx Q\) and some fixed atomic set \(\tilde{\mathcal{A}}_x\). Fix a probability tolerance \(\eta \in (0,1)\), and define the normalized total statistical dimension \(\Delta:=   d^{-1}\bigl( \sdim(\Desc(\tilde{\mathcal{A}}_x,\vct x_0)) + \sdim(\Desc(\mathcal{A}_y,\vct y_0))\bigr)\).  Then there is a scalar \(C>0\) that depends only on \(\eta\) such that
  \begin{align*}
  \Delta  &\le 1 - C/\sqrt{d} \implies \text{demixing can succeed with probability}\ge 1-\eta \\
 \Delta  &\ge 1 + C/\sqrt{d} \implies\text{demixing always fails with probability} \ge 1-\eta.
  \end{align*}
  By ``demixing can succeed,'' we mean that there exists a regularization parameter \(\lambda >0\) so that \((\vct x_0,\vct y_0)\) is an optimal point of~\eqref{eq:demix}.  ``Demixing always fails'' means that \((\vct x_0,\vct y_0)\) is not an optimal point of~\eqref{eq:demix} fails for \emph{any} parameter \(\lambda >0\).  
\end{theorem}
\begin{figure}[!t]
  \centering
    \includegraphics[width=0.4\columnwidth]{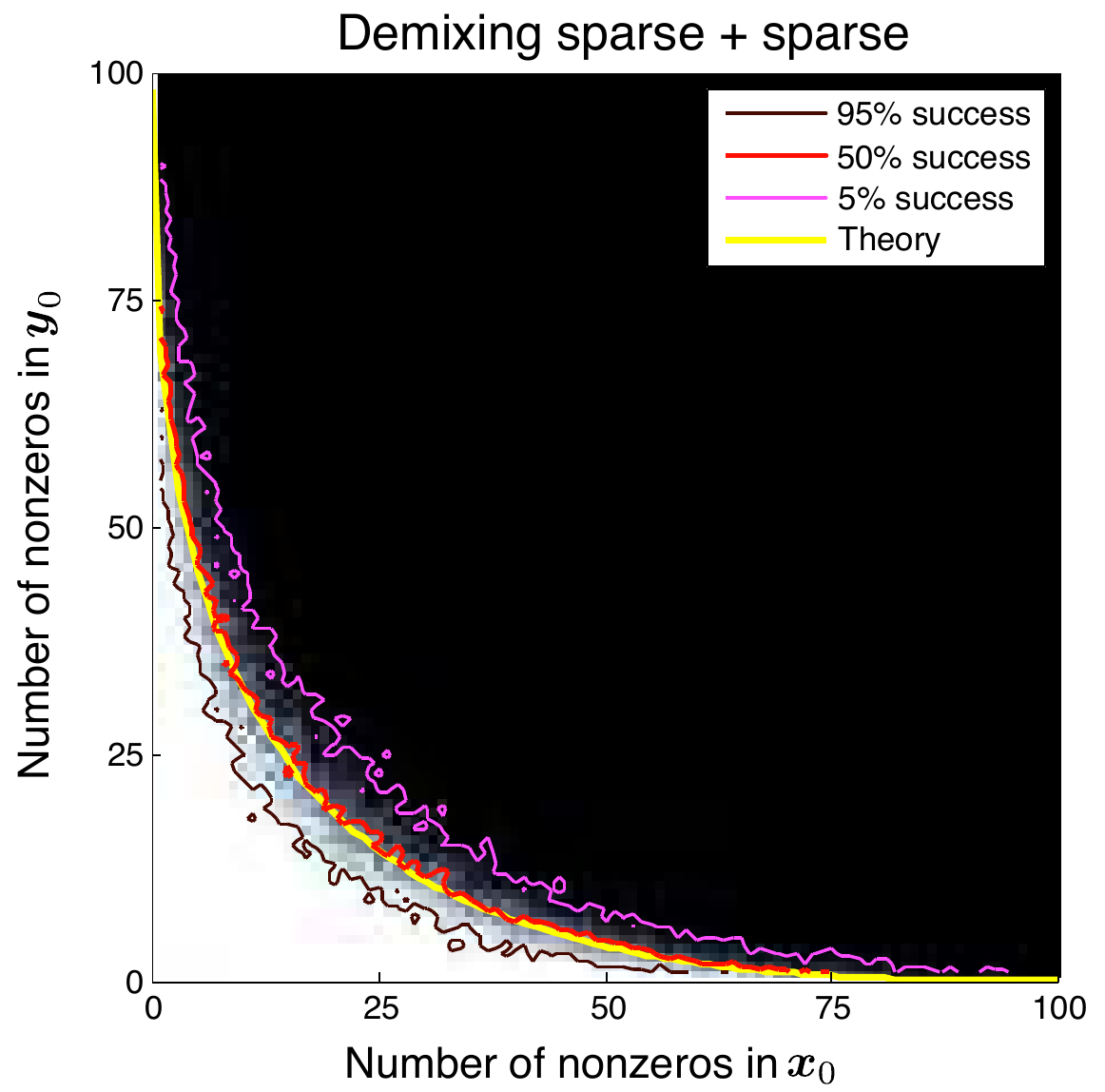}
  \caption{{ Phase transitions in demixing.} \it  Phase transition diagram for demixing two sparse signals using \(\ell_1\) minimization~\cite{McCTro:12,AmeLotMcC:13}. This experiment  replaces the DCT matrix \(\mtx D\) in~\eqref{eq:ell1} with a random rotation \(\mtx Q\).  The colormap shows the transition from pure success {\it (white)} to complete failure {\it (black)}. The 95\%, 50\%, and 5\% empirical success contours  {\it (tortuous curves)} appear above the theoretical phase transition curve {\it (yellow)} where \(\Delta = 1\). See~\cite{McCTro:12} for experimental details.}
  \label{fig:phaseT}
\vspace{-8mm}
\end{figure}

Theorem~\ref{thm:phase-trans} indicates that demixing exhibits a \emph{phase transition} as the total statistical dimension increases beyond the ambient dimension.  Indeed, if the total statistical dimension is slightly less than the ambient dimension, we can be confident that demixing will succeed, but if the total statistical dimension is slightly larger than the ambient dimension, then demixing is hopeless.   See Figure~\ref{fig:phaseT} for an example of the accuracy of this theory for the MCA model from the introduction when the DCT matrix \(\mtx D\) is replaced with a random rotation \(\mtx Q\).  The agreement between the empirical \(50\%\) success line and the curve where \(\Delta=1\) is remarkable.

This theory extends analogously to the compressive and multiple demixing models~\eqref{eq:gen1_demix} and~\eqref{eq:gen2_demix}. Under a similar incoherence model as above, compressive and multiple demixing are likely to succeed if and only if the total statistical dimension is slightly less than the number of (possibly compressed) measurements~\cite[Thm.~A]{McCTro:13}.  This fact lets us interpret the statistical dimension \(\sdim(\Desc(\mathcal{A},\vct x_0))\) as the degrees-of-freedom of the signal \(\vct x_0\) with respect to the atomic set \(\mathcal{A}\).  The message is clear: Incoherent demixing can succeed if and  only if the total dimension of the observation exceeds the total degrees-of-freedom of the constituent signals.

\makeatletter{}\section{Practical demixing algorithms}
\label{sec:numerics:-burn-it}
\emph{In theory}, many demixing problem instances of the form~\eqref{eq:demix} admit efficient numerical solutions. Indeed, if we can transform these problems into standard linear, cone, or semidefinite formulations, we can apply black-box interior point methods to obtain high-accuracy solutions in polynomial time~\cite{Nesterov2004}. \emph{In practice}, however, the computational burden of interior point methods makes these methods impracticable  as the dimension $d$  of the problem grows. 
Fortunately, a simple and effective iterative algorithm for computing approximate solutions to the demixing program~\eqref{eq:demix} and its extensions can be implemented with just a few lines of high-level code.  

\vspace{-5pt}
\subsection*{Splitting the work}
\label{sec:splitting-work}
The simplest and most popular method for iteratively solving demixing programs goes by the name  \emph{alternating direction method of multipliers} (ADMM).   The key object in this algorithm is the \emph{augmented Lagrangian} function \(L_\rho\) defined by
\begin{equation*}
  L_\rho(\vct x,\vct y,\vct w) := \|\vct x\|_{\mathcal{A}_x} + \lambda \|\vct y\|_{\mathcal{A}_y} + \langle \vct w, \vct x + \vct y - \vct z_0\rangle + \frac{1}{2\rho} \| \vct x + \vct y - \vct z_0\|^2,
\end{equation*}
where \(\langle \cdot,\cdot\rangle\) denotes the usual inner product between two vectors and \(\rho>0\) is a parameter that can be tuned to the problem.  Starting with arbitrary points \(\vct x^1,\vct y^1,\vct w^1 \in \R^d\), the ADMM method generates a sequence of points iteratively as 
\begin{equation}\label{eq:ADMM}
  \begin{cases}
    \vct x^{k+1} &= \argmin_{\vct x\in\R^d} L_{\rho}(\vct x, \vct y^k,\vct w^k)   \\
    \vct y^{k+1} &= \argmin_{\vct y\in \R^d} L_{\rho}(\vct x^{k+1},\vct y,\vct w^k) \\
    \vct w^{k+1} &= \vct w^k + (\vct x^{k+1} + \vct y^{k+1} -
    \vct z_0)/\rho.
  \end{cases}
\end{equation}
In other words, the \(\vct x\)- and \(\vct y\)-updates  iteratively minimize the Lagrangian over just \emph{one} parameter, leaving all others fixed.   The alternating minimization of \(L_\rho\)  gives the method its name.  Despite the simple updates, the  sequence \((\vct x^k,\vct y^k)\) of iterates generated in this manner converges to the minimizers \((\hvct x,\hvct y)\) of the demixing program~\eqref{eq:demix} under fairly general conditions~\cite{Combettes2005}.

The key to the efficiency of ADMM comes from the fact that the  updates are often easy to compute.  By completing the square, the \(\vct x\)- and \(\vct y\)-updates above amount to evaluating \emph{proximal operators} of the form
\begin{equation}\label{eq:prox}
  \vct x^{k+1} = \argmin_{\vct x\in \R^d} \|\vct x\|_{\mathcal{A}_x} + \frac{1}{2\rho}\|\vct u^k - \vct x\|^2 \qtq{and} \vct y^{k+1} = \argmin_{\vct y\in \R^d} \lambda\|\vct y \|_{\mathcal{A}_y} + \frac{1}{2\rho} \|\vct v^{k} - \vct y\|^2,
\end{equation}
where \(\vct u^{k} := \vct z_0 - \vct y^{k} - \rho \vct w^k\) and \(\vct v^k := \vct z_0 - \vct x^{k+1} - \rho \vct w^k\).    When solutions to the proximal minimizations~\eqref{eq:prox} are simple to compute, each  iteration of ADMM is highly efficient.

Fortunately, proximal operators are easy to compute for many atomic gauges.  For example, when the atomic gauge is the \(\ell_1\) norm, the proximal operator corresponds to soft-thresholding by~\(\rho\):
\begin{equation*}
  \argmin_{\vct x\in \R^d} \lone{\vct x} + \frac{1}{2\rho}\|\vct u - \vct x\|^2 = \mathrm{soft}(\vct u, \rho) =
  \begin{cases}
    u_i - \rho , & u_i > \rho , \\[-6pt]
    0, & |u_i| \le \rho , \\[-6pt]
    u_i + \rho , & u_i < \rho.
  \end{cases}
\end{equation*}
If we replace the \(\ell_1\) norm above with the Schatten-1 norm, then the corresponding proximal operator amounts to soft thresholding the singular values.    Numerous other explicit examples of proximal operations appear in~\cite[Sec.~2.6]{Combettes2005}.

Not all atomic gauges, however, have efficient proximal operations.  Even sets with finite number of atoms do not necessarily lead to more efficient proximal maps than sets with an infinite number of atoms. For instance, when the atomic set consists of rank-one matrices with unit Frobenius norm, we have an infinite set of atoms and yet the proximal map can be efficiently obtained via singular value thresholding. On the other hand, when the atomic set consists of rank-one matrices with binary $\pm 1$ entries, we have a finite set of atoms and yet the best-known algorithm for computing the proximal map  requires an intractable amount of computation.  

There is some hope, however, even for difficult gauges. Recent algebraic techniques for approximating atomic gauges  provide  computable  proximal operators in a relatively efficient manner, which opens the door to additional demixing algorithms for richer signal structures~\cite{chandrasekaran2012convex,bach2010structured}.

\subsection*{Extensions}
While the ADMM method is the prime candidate for solving problem \eqref{eq:demix}, it is not usually the best method for the extensions \eqref{eq:gen1_demix} or \eqref{eq:gen2_demix}. In the first case, if $\vct \Phi $ is a general linear operator, it creates a major computational bottleneck since we need an additional loop to solve the subproblems within the ADMM algorithm. In the latter case, ADMM even loses convergence guarantees  \cite{Chen2013b}. 

One possible way to handle both problems \eqref{eq:gen1_demix} and \eqref{eq:gen2_demix} is to use decomposition methods.
Roughly speaking, these methods decompose  problems \eqref{eq:gen1_demix} or \eqref{eq:gen2_demix} into smaller components and then solve the convex subproblem corresponding to each term simultaneously.  For example, we can use the decomposition method from  \cite{Chen1994}:
\begin{equation}\label{eq:prox_alm}
\begin{cases}
  \vct v^{k} &= \vct w^k + \rho (\vct{\Phi}(\vct x^{k} + \vct y^{k}) - \vct z_0) \\
  \vct x^{k+1} &= \argmin_{\vct x\in\R^d} \|\vct x\|_{\mathcal{A}_x} + \langle \vct v^{k},\vct{\Phi}\vct x\rangle + \frac{1}{2\rho}\|\vct{x} - \vct{x}^{k}\|^2_2\\
  \vct y^{k+1} &= \argmin_{\vct y\in \R^d} \lambda\|\vct y\|_{\mathcal{A}_y} + \langle\vct v^{k},\vct{\Phi}\vct y\rangle + \frac{1}{2\rho}\|\vct{y} - \vct{y}^{k}\|^2_2\\
  \vct w^{k+1} &= \vct w^k + \rho (\vct{\Phi}(\vct x^{k+1} + \vct y^{k+1}) - \vct z_0). 
\end{cases}
\end{equation}
When the parameter $\rho$ is chosen appropriately, the generated sequence $\{(\vct{x}^k, \vct{y}^k)\}$ in \eqref{eq:prox_alm} converges to the solution of \eqref{eq:gen1_demix}.  Since  the second and the third lines of \eqref{eq:prox_alm} are independent, it is even possible to solve them in parallel.  This scheme  easily extends to  demixing three or more signals~\eqref{eq:gen2_demix}.

Another practical method appears in~\cite{Necoara2008}. In essence, this approach combines a dual formulation, Nesterov's smoothing technique, and the fast gradient method \cite{Nesterov2004}. This technique works  both for problems \eqref{eq:gen1_demix} and \eqref{eq:gen2_demix}, and it possesses a rigorous $\mathcal{O}(1/k)$ convergence rate.

\makeatletter{}
\section{Examples}\label{sec:examples}

The ideas above apply to a large number of examples.  Here, we highlight some recent applications of convex demixing in signal  processing.    The first example, texture inpainting, uses a low-rank and sparse decomposition  to discover and repair axis-aligned texture in images.  The second example explores an application of demixing to direction-of-arrival estimation, where we demix  a source covariance from a noise covariance to improve beamforming. 
\vspace{-2mm}
\subsection*{Texture inpainting}
  Many natural and man-made images include highly regular textures. These repeated patterns, when aligned with the image frame,  tend to have very low rank.  Of course, rarely does a natural image consist solely of a texture. Often, though, a background texture is \emph{sparsely} occluded by a untextured component. By modeling the occlusion as an additive error, we can use convex demixing  to solve for the underlying texture and extract the occlusion~\cite{liang2012repairing}. 

\begin{figure}
\centering
\includegraphics[width=0.8\columnwidth]{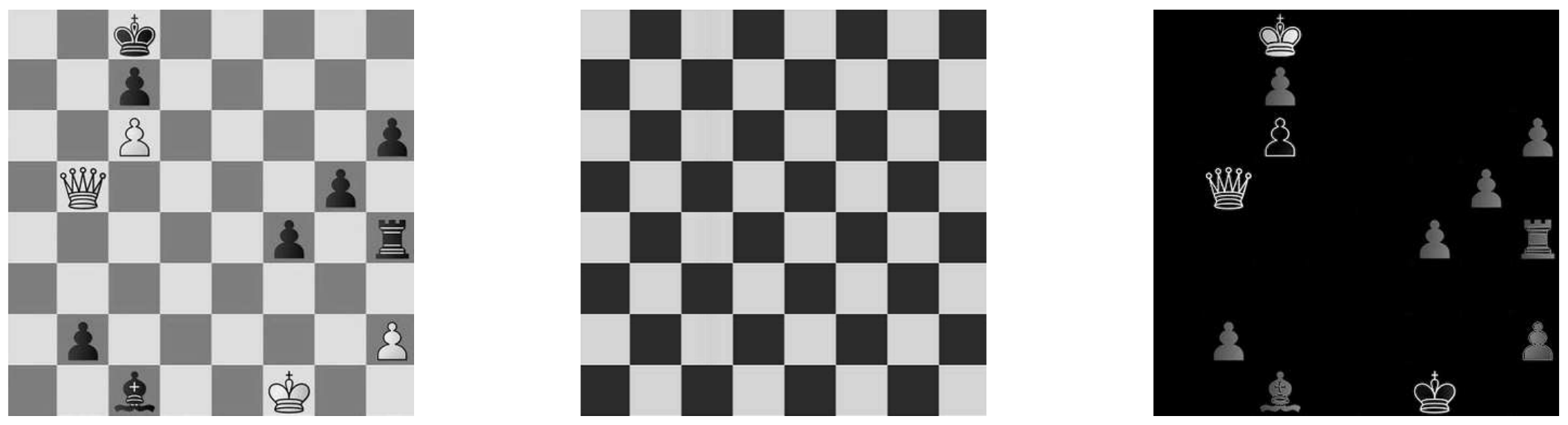}
\caption{\label{fig:rpca}{Texture inpainting (White to move, checkmate in $2$).} \it The rank-sparsity decomposition~\eqref{eq:texture-repair} perfectly separates the chessboard from the pieces. {\it (Left)} Original image. {\it (Center)} Low-rank component. {\it (Right)} Sparse component.
}
\vspace{-8mm}
\end{figure}

In this model, we treat the observed digital image \(\mtx Z_0 \in \R^{m\times n}\) as a matrix formed by the sum \(\mtx Z_0 = \mtx X_0 + \mtx Y_0\), where the textured component \(\mtx X_0\) has low rank and \(\vct Y_0\) is a sparse corruption or occlusion.  The natural demixing program in this setting is the rank-sparsity decomposition~\cite{ChaSanPar:11,CanLiMa:11}:
\begin{equation}\label{eq:texture-repair}
  [\hat{\vct X}, \hat{\vct Y}] = \argmin_{\vct X ,\mtx Y \in \R^{m\times n}}  \norm{\mtx X}_{S_1} + \lambda \norm{ \mtx Y}_1 \subjectto \vct X + \vct Y = \vct Z_0,
\end{equation}
This unsupervised texture-repair method exhibits state-of-the-art performance, exceeding even the quality of a supervised procedure built in to Adobe Photoshop\textregistered\ on some images~\cite{liang2012repairing}.  When applied, for example, to an image of a chessboard, the method flawlessly recovers the checkerboard from the pieces (Figure~\ref{fig:rpca}).

\vspace{-3mm}
\subsection*{Direction-of-arrival estimation}
\label{sec:direct-arriv-estim}

We describe a convex demixing program for direction-of-arrival (DOA) estimation.  In DOA  estimation, we use  an array of \(n\) sensors to determine the bearing of multiple sources in wireless communications~\cite{saunderson2012diagonal}.  When the sources are independent, the joint covariance matrix \(\vct Z_0\) of all of the signals takes the form
$
\mtx  Z_0 =  \mtx A_0 \mtx A_0^\transp + \mtx Y_0
$
in \emph{expectation}, where the column space of the \(n\times r\) matrix \(\mtx A_0\) encodes the bearing information from \(r\) sources, and \(\mtx Y_0\) is the covariance matrix of the noise at the sensors.  

When the number  of sources \(r\) is much smaller than the number  of sensors \(n\), the matrix \(\mtx X_0 := \mtx A_0 \mtx A_0^\transp\) is positive semidefinite and has low rank.  Moreover, when the sensor noise is uncorrelated, the matrix \(\mtx Y_0\) is diagonal.  Using the atomic gauge recipe from above, we can demix \(\mtx X_0\) and \(\mtx Y_0\) from the empirical covariance matrix \(\vct Z_0\) by setting
\begin{equation}\label{eq: bearing}
  [\hat{\mtx X},\hat{\mtx Y}, \hat{\mtx E}] = \argmin_{\mtx X,\mtx Y \in \R^{n\times n}} \quad \norm{\mtx X}_{S_1^+} + \norm{\mtx Y}_{\mathrm{diag}} + \lambda \norm{\mtx E}^2_{\mathrm{Fro}} \subjectto \mtx X + \mtx Y + \mtx E = \mtx Z_0,
\end{equation}
where $\mtx E$ absorbs the deviations in the expectation model due to the finite sample size.   Here, \(\norm{\cdot}_{S_1^+}\) is the atomic gauge generated by positive semidefinite rank-one matrices, which is equal to the trace for positive semidefinite matrices, but returns \(+\infty\) when its argument has a negative eigenvalue. Similarly, the gauge  \(\norm{\cdot}_{\mathrm{diag}}\) is the atomic gauge generated by the set of all diagonal matrices, and so it is equal to zero on diagonal matrices but \(+\infty\) otherwise.   The norm \(\norm{\cdot}_\mathrm{Fro}\) is the usual Frobenius norm on a matrix. The results of~\cite{saunderson2012diagonal} relate the success of a similar problem to the geometric problem of ellipsoid fitting, and show that under some incoherence conditions convex demixing succeeds. 

In DOA estimation, the source covariance matrix plays a key role in estimating the source directions \cite{VanTrees}. For instance, the multiple signal classification (MUSIC) algorithm exploits the nullspace of the source covariance matrix to localize the sources. In the presence of white additive Gaussian noise, the empirical covariance estimate becomes corrupted,  deteriorating the bearing estimates generated by MUSIC. 

\begin{figure}[!t]
  \centering
\begin{tabular}{cc}
  \includegraphics[width=0.32\columnwidth]{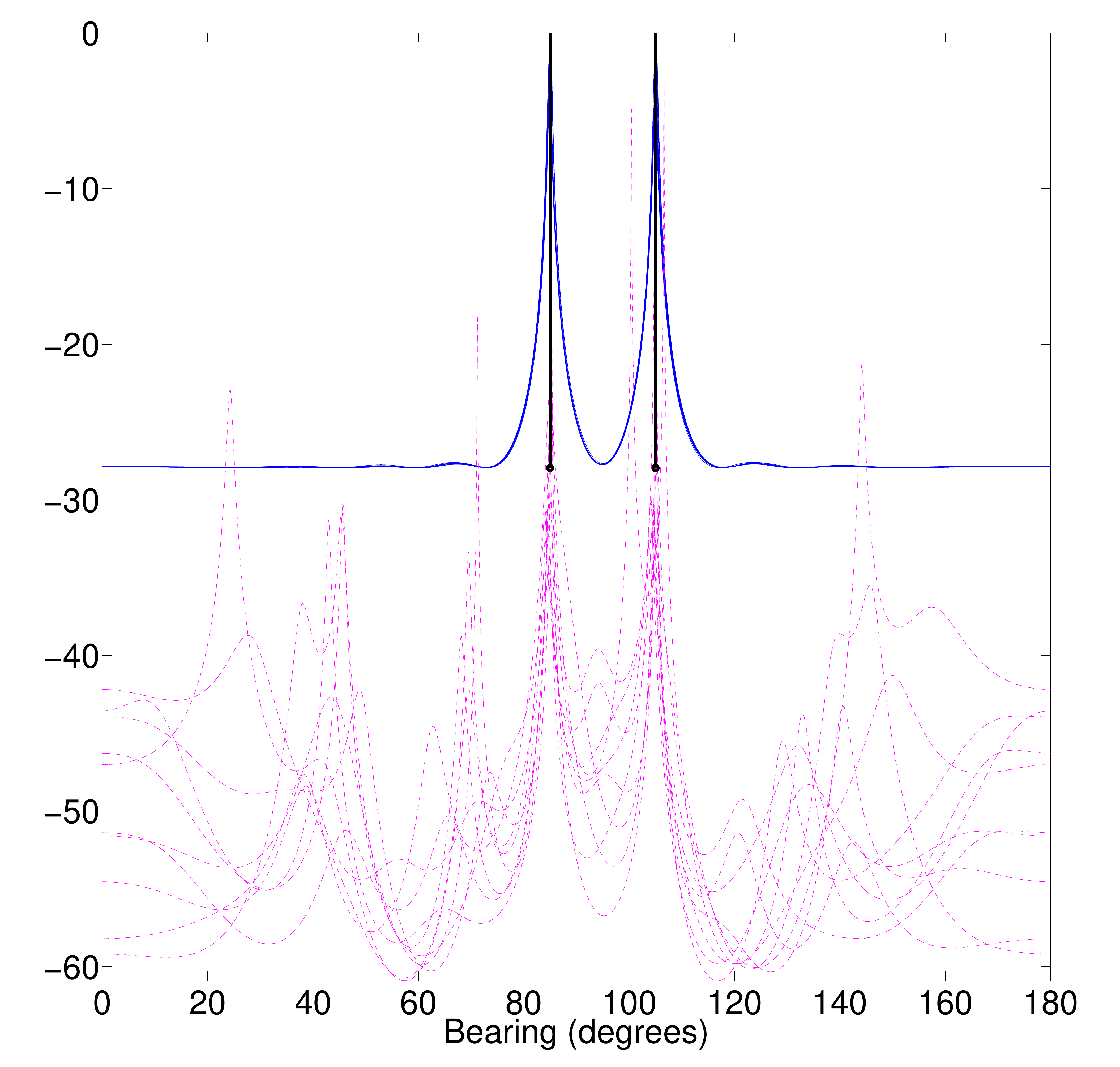} &
  \includegraphics[width=0.32\columnwidth]{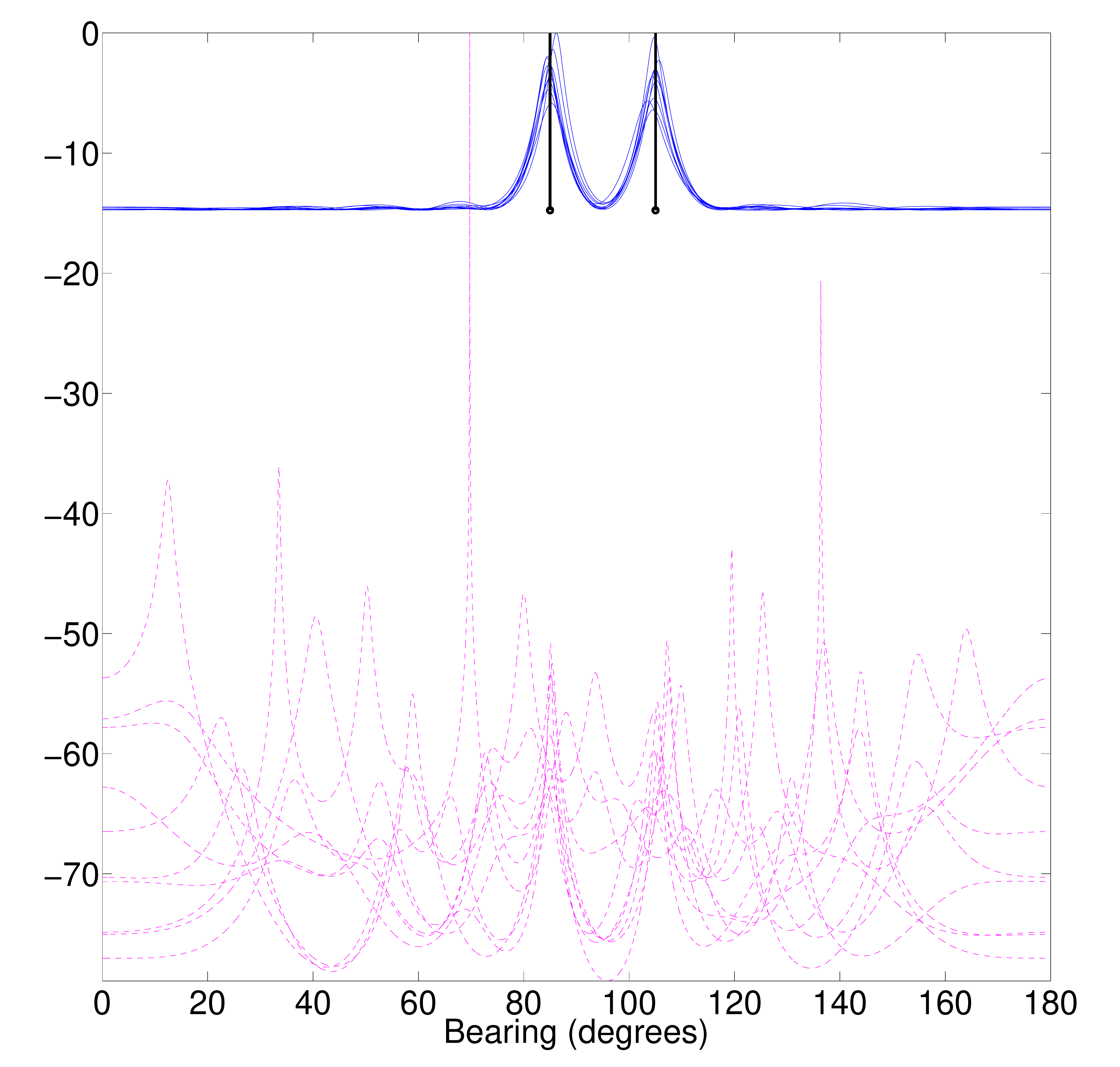} \\
     MUSIC pseudospectrum at $5$dB & MUSIC pseudospectrum at $-5$dB 
\end{tabular}  
  \caption{Enhancing DOA estimation. {\it The MUSIC pseudospectrum based on the demixed estimate $\hvct X$ (solid blue lines) from~\eqref{eq: bearing} is significantly more informative for the source bearings than the MUSIC pseudospectrum based on the raw covariance $\mtx Z_{0}$ (dashed magenta lines).}}
\label{fig:res}
\vspace{-6mm}
\end{figure}

Figure \ref{fig:res} shows how the demixing procedure~\eqref{eq: bearing} can significantly boost the performance of MUSIC under additive noise. In this experiment, we generate an array data for $r=2$ sources and $n=10$ sensors with signal-to-noise ratios of $5$dB and $-5$dB. We simulate the data and compute the empirical covariance matrix $\mtx Z_0$.  Then we estimate the source covariance \(\vct X_0\) using the demixed output \(\hvct X\) of~\eqref{eq: bearing}. We compare the performance of MUSIC with given the raw empirical covariance \(\mtx Z_0\) and the demixed estimate \(\hvct X\).

At $5$dB SNR, about one-third of the DOA estimates of the MUSIC algorithm with $\mtx Z_0$ are more than three degrees off of the true bearings. At $-5$dB, MUSIC's  performance on the raw covariance is even worse: \(90\%\) of the estimated bearings are off by three degrees or more. In  contrast, the MUSIC algorithm using the demixed estimate $\hvct X$ provides consistently accurate  bearing estimates.

\makeatletter{}
\section{Horizons: Nonlinear separation}
\label{sec:new-direct-nonl}
We conclude our demixing tutorial with some promising directions  for the future. In many applications, the constituent signals are tangled together in a  \emph{nonlinear} fashion \cite{ahmed2012blind,BitReRec:12}.  While this situation would seem to rule out the linear superposition model  considered above, we can  leverage the same convex optimization tools to obtain demixing guarantees and often return to a linear model using a  technique called \emph{semidefinite relaxation}.  

We describe the basic idea behind this maneuver with a concrete application: \emph{blind deconvolution}. Convolved signals appear  frequently in communications due, for example, to multipath channel effects. When the channel is known, removing the channel effects is a difficult but well-understood linear inverse problem.  With blind deconvolution, however, we see only the convolved signal
\begin{math}
  \vct z_0 = \vct x_0 * \vct y_0
\end{math}
from which we must determine both the channel \(\vct x_0 \in \R^m\) and the source \(\vct y_0 \in \R^d\). 

While the convolution \(\vct x_0 *\vct y_0\) involves nonlinear interactions between \(\vct x_0\) and \(\vct y_0\), the convolution is in fact \emph{linear} in the matrix  formed by the outer product \(\vct x_0 \vct y_0^\transp\).   In other words, there is a linear operator \(\mathcal{C}\colon \R^{m\times d}\to \R^{m+d}\) such that
\vspace{-3mm}
\begin{equation*}
 \vct z_0 = \mathcal{C}\bigl(\mtx X_0\bigr) \qtq{where} \mtx X_0 := \vct x_0 \vct y_0^\transp.\vspace{-3mm}
\end{equation*}
The matrix \(\vct X_0\) has rank one by definition, so it is natural use the Schatten 1-norm to search for low-rank matrices that generate the observed signal:
\begin{equation*}
  \hat{\mtx X} = \argmin_{\mtx X \in\R^{m\times d}}\quad \norm{\mtx X}_{S_1} \subjectto  \vct z_0 = \mathcal{C}(\vct X).
\end{equation*}
This is the basic idea behind the convex approach to blind deconvolution of~\cite{ahmed2012blind}.

The implications of the non-linear demixing example above are far reaching.  There are large classes of signal and mixing models that support efficient, provable, and stable demixing. Viewing different demixing problems within a common framework of convex optimization, we can leverage decades of research in various diverse disciplines from applied mathematics to signal processing, and from theoretical computer science to statistics. We expect that the diversity of convex demixing models and geometric tools will also inspire the development of new kinds of scalable optimization algorithms that handle non-conventional cost functions along with atomic gauges \cite{TranDinh2013c}.

{
\setstretch{1}

\bibliographystyle{IEEEtran}

}

\end{document}